%% file: 00-main.tex
\documentclass[conference,a4paper]{IEEEtran}
\IEEEoverridecommandlockouts

\usepackage{amsmath,amssymb,amsfonts}
\usepackage{csquotes}
\usepackage{algorithmic}
\usepackage{graphicx}
\usepackage{textcomp}
\usepackage{xcolor}
\usepackage{soul}
\usepackage{balance}
\usepackage{siunitx}

\def\BibTeX{{\rm B\kern-.05em{\sc i\kern-.025em b}\kern-.08em
    T\kern-.1667em\lower.7ex\hbox{E}\kern-.125emX}}
    
\usepackage[style=numeric,
  isbn=false,
  doi=false,
  sorting=none,
  url=false,
  bibencoding=utf8,
  backend=biber,
  style=numeric,
  mincitenames=1, maxcitenames=1]{biblatex}

\usepackage[english]{babel}
\usepackage{booktabs}
\usepackage{multirow}

\makeatletter
 \let\old@ps@headings\ps@headings
 \let\old@ps@IEEEtitlepagestyle\ps@IEEEtitlepagestyle
 \def\confheader#1{%
 \def\ps@headings{%
 \old@ps@headings%
 \def\@oddhead{\strut\hfill#1\hfill\strut}%
 \def\@evenhead{\strut\hfill#1\hfill\strut}%
 }%
 \def\ps@IEEEtitlepagestyle{%
 \old@ps@IEEEtitlepagestyle%
 \def\@oddhead{\strut\hfill#1\hfill\strut}%
 \def\@evenhead{\strut\hfill#1\hfill\strut}%
 \def\@oddfoot{\mycp}%
 \def\@evenfoot{\mycp}
 }%
 \ps@headings%
 }
\makeatother

\confheader{2021 International Conference on Indoor Positioning and Indoor Navigation (IPIN), 29 Nov. -- 2 Dec. 2021, Lloret de Mar, Spain}

\newcommand{\mycp}{978-1-6654-0402-0/21/\$31.00~\copyright~2021~IEEE\hfill}

\renewcommand{\mycp}{}
\confheader{}

\begin{document}

\title{New trends in indoor positioning based on WiFi and machine learning: A systematic review
% \thanks{Este trabajo ha sido parcialmente co-financiado por las siguientes instituciones. }
% \thanks{This work has been partially co-financed by the Ministry of Economy and Competitiveness with the projects REPNIN+ (TEC2017-90808-REDT) e~INSIGNIA (Torres-Quevedo program, PTQ2018-009981).}
}
\author{
\IEEEauthorblockN{
Vladimir Bellavista-Parent\IEEEauthorrefmark{1}, 
Joaquín Torres-Sospedra\IEEEauthorrefmark{2}, 
and Antoni Perez-Navarro\IEEEauthorrefmark{1}\textsuperscript{,}\IEEEauthorrefmark{3}
}
\IEEEauthorblockA{\IEEEauthorrefmark{1}\textit{Internet Interdisciplinary Institute (IN3)}, Universitat Oberta de Catalunya, Castelldefels, Spain}
%\IEEEauthorblockA{\IEEEauthorrefmark{2}\textit{Institute of New Imaging Technologies}, \textit{Universitat Jaume I}, Castellón, Spain}
\IEEEauthorblockA{\IEEEauthorrefmark{2}\textit{UBIK Geospatial Solutions S.L.}, Castellón, Spain}
\IEEEauthorblockA{\IEEEauthorrefmark{3}\textit{Faculty of Computer Sciences, Multimedia and Telecommunication}, Universitat Oberta de Catalunya, Barcelona, Spain}
\texttt{\url{vbellavista@uoc.edu}} -- \texttt{\url{torres@ubikgs.com}} -- \texttt{\url{aperezn@uoc.edu}}}%CORREOS

% \author{%
% \IEEEauthorblockN{%
% Lista de Autores\\Completa\\ {\ }\\%
% }
% \IEEEauthorblockA{\textit{Institución},\\ Ciudad, País}
% }

% \and
% \and
% \and
% \and
% \and

\maketitle

\begin{abstract}
Currently there is no standard indoor positioning system, similar to outdoor GPS. However, WiFi signals have been used in a large number of proposals to achieve the above positioning, many of which use machine learning to do so. But what are the most commonly used techniques in machine learning? What accuracy do they achieve? Where have they been tested? This article presents a systematic review of works between 2019 and 2021 that use WiFi as the signal for positioning and machine learning models to estimate indoor position. 64 papers have been identified as relevant, which have been systematically analyzed for a better understanding of the current situation in different aspects. The results show that indoor positioning based on WiFi trends use neural network-based models, evaluated in empirical experiments. Despite this, many works still conduct an assessment in small areas, which can influence the goodness of the results presented.
\end{abstract}


\input{01-introduction}

\input{02-relatedwork}
\input{03-method}
\input{04-results}

\input{05-discussion}
\input{06-conclusions}

\input{07-acknowledgments}

\renewcommand*{\UrlFont}{\rmfamily}
\printbibliography
\balance
\end{document}

%% file: 01-introduction.tex
\section{Introduction}
Nowadays the outdoor location is resolved in most of the situations, thanks to the GNSS \textit{(Global Navigation Satellite Systems)}, among which, the best known are GPS, GLONASS and GALILEO. They are widely used systems all over the planet and we only need one receiver (e.g., a smartphone) to get the position in which we are.
The problem arises in indoor spaces, as all these systems stop working due to the low capacity of penetration of the signal into the buildings.

The field of indoor positioning has been advancing in different ways for some years and with different technologies: inertial~\cite{Jimenez2010135}, Bluetooth~\cite{Faragher2014201}, WiFi~\cite{1356988}, Ultrasounds~\cite{s18010089}, Visible Light~\cite{Yoshino2008}, among others; and even combining them with each other, what we know as sensor fusion \cite{7275538,s20061717}. 
In recent years, due to advances in machine learning, many of the work is based on applying machine learning algorithms. In addition, the emergence of new technologies, like 5G~\cite{8757118} or  WiFi mmWave~\cite{RN8098}, favour alternatives for indoor positioning.

Nevertheless, the most popular indoor positioning mechanisms are based on WiFi. However, the large number of articles published in this topic in recent years requires a compilation of them and classify them so that they can be useful for future research. This work presents a systematic review of the use of machine learning algorithms for indoor positioning based on WiFi signals in different papers.  Specifically, publications since January 2019 are analyzed, of which detailed information is provided on the algorithms used, as well as on different aspects that have been considered relevant such as: type of work performed (experimental/simulated), size of the test field, number of access points (APs), number of reference points for the radio map, results obtained, evaluation metrics used, type of signal, type of technology used and use of rooms or not for the experiments. 

The remainder of this work is organised as follows. Section II briefly reviews the literature related to reviews about indoor positioning. Section III describes the methodology followed in this article. Section IV presents the results with a detailed table. Results are analyzed in Section V. Finally, Section VI gives the conclusion.

%% file: 02-relatedwork.tex
\section{Related Work}
A variety of machine learning approaches have been proposed for indoor localization with WiFi and Machine Learning. Therefore, there are other reviews related to this topic but they are based on other research questions and have different objectives. For example, \cite{Pascacio2021} does an exhaustive analysis of the different indoor positioning articles but focused on articles using collaborative positioning techniques. These techniques are based on the exchange of information between different users and/or devices to improve the overall positioning of the system. These methods have the advantage that requires smaller infrastructure than other methods to create and maintain an indoor positioning system. Their main disadvantage is its greatest computational need, since positioning depends on the calculation of data from multiple sources (users and/or devices).

In \cite{KimGeok2020},  we can see a tune-up of all the wireless technologies used in the field of indoor positioning, doing an in-depth analysis of all of them. Thus, they consider works based on the received signal (RSSI or CSI), and those that use data such as Time of Flight (ToF), Angle of Arrival (AoA) or Phase of Arrival (PoA). They also review the different algorithms used to achieve positioning, such as fingerprinting, multilateration or triangulation.  Finally, the work classifies the most used Machine Learning algorithms and the methods used to filter received signals. However, it is a review of the technologies used and does not analyze the contribution of every paper individually.

A similar paper, \cite{Obeidat2021}, explains other systems and does not focus only on works based on radio signals. It includes any type of wireless signal, even those based on optical or magnetic solutions. The authors also analyze the different algorithms and techniques used to achieve positioning. However, they also do not go into sorting the paper individually.

Finally, \cite{Alhomayani2020} makes an exhaustive review of all the deep learning algorithms used in any study that uses fingerprint as a method for positioning. In this review there is also a compilation of the most used public WiFi radio maps and a brief analysis of each one of them. As in the rest of the reviews mentioned above, it focuses more on the analysis of the different elements involved in an indoor positioning job than on analyzing the individual items.

%% file: 03-method.tex
\section{Methodology}
\label{sec:metodos}
The methodology that has been followed in this work to carry out the systematic review is based on the PRISMA~\cite{Page2021} guidelines. Thus, a set of questions has been first established to set the objective of the review. The next step has been to define the different search queries to be executed in the digital libraries in which the articles of interest for this work reside. Next, a set of inclusion and exclusion criteria has been set which has been expanded to a full final version that has resulted in the final selection of articles that are part of this work.  The following are the research questions:
\renewcommand{\theenumi}{\textbf{RQ\arabic{enumi}}}
\begin{enumerate}
\item Which machine learning algorithms get the best results in WiFi-based indoor positioning?
\item What kind of WiFi signal parameter gets the best results?
\item What are the most commonly used metrics in indoor positioning studies?
\item Are there substantial differences between simulated and experimental studies?
\item Which public radio signal maps are the most commonly used in simulation?

\end{enumerate}

To perform queries, \emph{Web of Science} and \emph{Scopus} database has been chosen because they are reliable sources with sufficient content. %
%
%\emph{Queries} performed are displayed in Figures \ref{fig:queryscopus} and \ref{fig:querywos}.
%
Figures \ref{fig:queryscopus} and \ref{fig:querywos} introduce the queries we have used to retrieve the scientific papers in those two datasets.%from the databases \emph{Scopus} and \emph{Web of Science} respectively.

\begin{figure}[!htb]
\scriptsize{\begin{verbatim}
TITLE-ABS-KEY (
    (wifi OR wi-fi OR wlan OR "802.11" 
        OR wireless ) AND  indoor AND 
    (posi* OR loca* OR  nav* OR track*) AND
    (deep* OR machine* OR *nn OR svm OR svr 
           OR neural OR convolutional 
           OR "decision tree" OR "boosting" 
           OR "bagging" OR "particle filter"))
AND (LIMIT-TO (PUBYEAR,2021) OR LIMIT-TO (PUBYEAR,2020) OR 
     LIMIT-TO (PUBYEAR,2019))
\end{verbatim}}
    \caption{Query for \emph{Scopus}}
    \label{fig:queryscopus}
\end{figure}
%\textbf{Scopus}
%\texttt{\emph{TITLE-ABS-KEY ((5g OR wifi OR wi-fi OR wlan OR "802.11" OR wireless ) AND indoor AND (posi* OR loca* OR  nav* OR track*) AND (deep* OR machine* OR *nn OR svm OR svr OR neural OR convolutional OR "decision tree" OR "boosting" OR "bagging" OR "particle filter")) AND (LIMIT-TO (PUBYEAR , 2021) OR LIMIT-TO (PUBYEAR , 2020) OR LIMIT-TO (PUBYEAR , 2019) OR LIMIT-TO (PUBYEAR , 2018) OR LIMIT-TO (PUBYEAR , 2017) OR LIMIT-TO (PUBYEAR , 2016) OR LIMIT-TO (PUBYEAR , 2015))}}

\begin{figure}[!htb]
\scriptsize{\begin{verbatim}
(AB=((wifi OR wi-fi OR wlan OR 802.11 OR wireless) AND 
    indoor AND (posi* OR loca* OR nav* OR track*) AND
    (deep* OR machine* OR knn OR k-nn OR svm OR svr 
           OR  neural OR convolutional OR “decision tree” 
           OR boosting OR bagging OR “particle filter”))
 OR TI=(wifi OR wi-fi OR wlan OR 802.11 OR wireless) AND 
    indoor AND (posi* OR loca* OR nav* OR track*) AND 
    (deep* OR machine* OR knn OR k-nn OR svm OR svr 
           OR neural OR convolutional OR “decision tree”
           OR boosting OR bagging OR “particle filter”))
 OR AK=((wifi OR wi-fi OR wlan OR 802.11 OR wireless) AND 
    indoor AND (posi* OR loca* OR nav* OR track*) AND 
    (deep* OR machine* OR knn OR k-nn OR svm OR svr 
           OR neural OR convolutional OR “decision tree” 
           OR boosting OR bagging OR “particle filter”)))
 AND PY=(2019-2021)
\end{verbatim}}
    \caption{Query for \emph{Web of Science}}
    \label{fig:querywos}
\end{figure}
%\textbf {Web Of Science}
%\emph{(AB=((5g OR wifi OR wi-fi OR wlan OR 802.11 OR wireless) AND indoor  AND (posi* OR loca* OR nav* OR track*) AND (deep* OR machine* OR knn OR k-nn OR svm OR svr OR neural OR convolutional OR “decision tree” OR boosting OR bagging OR “particle filter”)) OR TI=((5g OR wifi OR wi-fi OR wlan OR 802.11 OR wireless) AND indoor AND (posi* OR loca* OR nav* OR track*) AND (deep* OR machine* OR knn OR k-nn OR svm OR svr OR neural OR convolutional OR “decision tree” OR boosting OR bagging OR “particle filter”))OR AK=((5g OR wifi OR wi-fi OR wlan OR 802.11 OR wireless) AND indoor AND (posi* OR loca* OR nav* OR track*) AND (deep* OR machine* OR knn OR k-nn OR svm OR svr OR neural OR convolutional OR “decision tree” OR boosting OR bagging OR “particle filter”))) AND PY=(2015-2021)}

The inclusion criteria that selected papers must satisfy are:

\renewcommand{\theenumi}{\textbf{IC\arabic{enumi}}}
\begin{enumerate}
\item Written in English
\item Coming from a conference or journal article
\item Dealing with WiFi-based positioning
\item Positioning with Machine Learning algorithms
\item Published between 2019 to 2021
\end{enumerate}

\begin{figure}[!b]
\centering
\includegraphics[width=0.8\columnwidth]{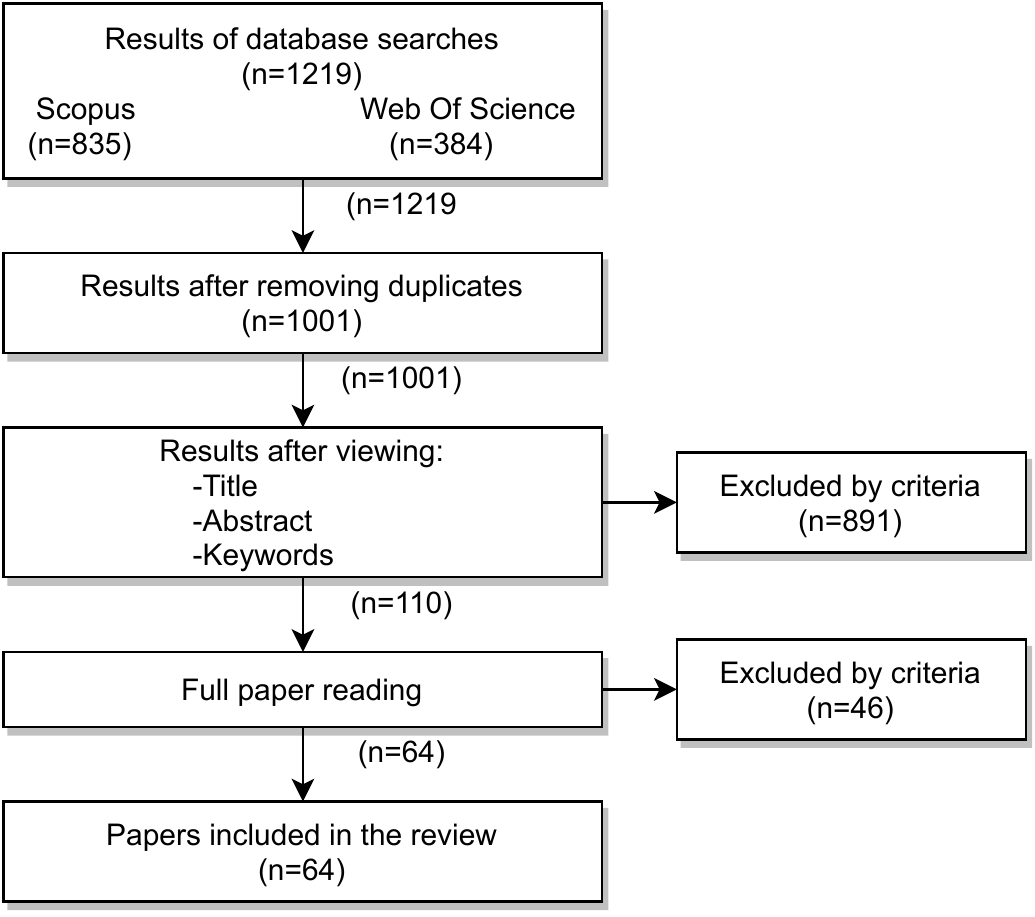}\\{\ }\\
\caption{PRISMA flow diagram}
\label{fig:foo}
\end{figure}

\input{Tables/table_papers}

The exclusion criteria are:

\renewcommand{\theenumi}{\textbf{EC\arabic{enumi}}}
\begin{enumerate}
\item Workshops and book chapters
\item Positioning that's not 100\% WiFi
\item Positioning that has some of the work outdoors
\item Positioning based on classic multilateration (TOA/TDOA/AOA/POA..)
\item Positioning based on Sensor Fusion, as we seek only to analyze the behavior of algorithms in WiFi signals
\item Positioning that uses KNN-based algorithm or Particle Filter, as it is not considered Machine Learning
\end{enumerate}

%\subsection{Proceso de selección}

Once all the results have been obtained, duplicates from the two datasets, \emph{Web of Science} and \emph{Scopus}, have been removed. With the resulting articles, a first analysis of the title and abstract of each of them has been carried out to rule out those who failed to meet the criteria of inclusion or met the exclusion criteria. Finally, a full reading has been made of the included articles to verify whether they met the inclusion criteria and analyse the elements that answer the research questions. %El diagrama completo puede verse en la Figura~\ref{fig:foo}.

%\begin{figure}[!ht]
%\centering
%\includegraphics[width=0.8\columnwidth]{diagramaPRISMA.pdf}
%\caption{Diagrama de flujo PRISMA}
%\label{fig:foo}
%\end{figure}

%% file: Tables/table_papers.tex
\begin{table*}
 \caption{Summary of revised articles}
 \label{tabla:review}
\resizebox{\textwidth}{!}{%
 \tabcolsep 2.925pt
%\begin{tabular}{c c c c c c c c c c c c c c p{2.5cm} p{4cm} p{4cm} p{4cm} } 
 \begin{tabular}{c c c c c c c c 
 c c c c c} 
 %\midrule%\hline
 \toprule
 \bf{art} & \bf{year} & \hbox to 1.5 cm{\hss\bf{WiFi}\hss} & \bf{est} & \bf{AP} & \bf{rPoint} & \bf{fMap} &  \bf{fmRoom} & \hbox to 2.5 cm{\hss\bf{mAlg}\hss}& \hbox to 3 cm{\hss\bf{sAlg}\hss} & \hbox to 1 cm{\hss\bf{mError}\hss} & \hbox to 6 cm{\hss\bf{oError}\hss} & \bf{sType} \\  
 \midrule

\input{Tables/contents1}

\bottomrule
\end{tabular}}
\begin{tabular}{p{1cm}p{7cm}p{1cm}p{5cm}}
\textbf{art}    & Article                                           &\textbf{fmRoom}  & Rooms used in exp/sim                  \\
\textbf{WiFi}   & WiFi type                                         &\textbf{mAlg}    & Main algorithm used              \\
\textbf{est}    & Experimental or simulated study                   &\textbf{sAlg}    & Other algorithms used in the study              \\
\textbf{AP}     & APs used                                          &\textbf{mError}  & Mean Error                \\
\textbf{rPoint} & Reference Points used in offline phase            &\textbf{oError}  & Other metrics reported in the study               \\
\textbf{fMap}   & Size of experimental room or radio-map used       &\textbf{sType}   & Signal type used                                \\
&\\

\end{tabular}
\end{table*}

\addtocounter{table}{-1}

\begin{table*}
 \caption{Summary of revised articles (Continuation)}
 \label{tabla:review2}
\resizebox{\textwidth}{!}{%
 \tabcolsep 2.925pt
 %\begin{tabular}{c c c c c c c c c c c c c c p{2.5cm} p{4cm} p{4cm} p{4cm} } 
 \begin{tabular}{c c c c c c c c 
 c c c c c} 
 %\midrule%\hline
 \toprule
\bf{art} & \bf{year} & \hbox to 1.5 cm{\hss\bf{WiFi}\hss} & \bf{est} & \bf{AP} & \bf{rPoint} & \bf{fMap} &  \bf{fmRoom} & \hbox to 2.5 cm{\hss\bf{mAlg}\hss}& \hbox to 3 cm{\hss\bf{sAlg}\hss} & \hbox to 1 cm{\hss\bf{mError}\hss} & \hbox to 6 cm{\hss\bf{oError}\hss} & \bf{sType} \\   
 \midrule

\input{Tables/contents2}
\input{Tables/contents3}

\bottomrule
\end{tabular}}
\begin{tabular}{p{1cm}p{7cm}p{1cm}p{5cm}}
\textbf{art}    & Article                                           &\textbf{fmRoom}  & Rooms used in exp/sim                  \\
\textbf{WiFi}   & WiFi type                                         &\textbf{mAlg}    & Main algorithm used              \\
\textbf{est}    & Experimental or simulated study                   &\textbf{sAlg}    & Other algorithms used in the study              \\
\textbf{AP}     & APs used                                          &\textbf{mError}  & Mean Error                \\
\textbf{rPoint} & Reference Points used in offline phase            &\textbf{oError}  & Other metrics reported in the study               \\
\textbf{fMap}   & Size of experimental room or radio-map used       &\textbf{sType}   & Signal type used                                \\

\end{tabular}
\end{table*}

%% file: Tables/contents1.tex
\multirow{3}[4]{*}{\cite{Dou2021}}&	\multirow{3}[4]{*}{2021}&	WiFi&	sim&	$168$&	$927$&	IPIN2016 Tutorial&	N&	DRL&	&	\SI{0,244}{\meter}&	&	RSSI	\\\cmidrule(lr){3-13}
&	&	WiFi&	sim&	$589$&	$1840$&	UTSIndoorLoc&	Y&	DRL&	&	\SI{0,219}{\meter}&	& RSSI		\\\cmidrule(lr){3-13}
&	&	WiFi&	sim&	$520$&	$993$&	UJIIndoorLoc&	Y&	DRL&	&	\SI{0,197}{\meter}&	& RSSI		\\\midrule
\multirow{2}[2]{*}{\cite{Roy2021}}&	\multirow{2}[2]{*}{2021}&	WiFi&	sim&	$96$&	$80$&	JUIndoorLoc&	Y&	BayesNet&	Dempster–Shafer&	&	Accuracy  =  80\% between \SI{3}{\meter} and \SI{3.6}{\meter}&	RSSI	\\\cmidrule(lr){3-13}
&	&	WiFi&	sim&	$520$&	$993$&	UJIIndoorLoc&	Y&	BayesNet& Dempster–Shafer	&	&	Accuracy  =  98\% in \SI{2}{\meter}&	RSSI	\\\midrule
\cite{Zhang2020}&	2020&	WiFi&	exp&	$3$&	$21$&	\SI{45}{\square\meter}&	Y&	CNN&	&	\SI{1,2669}{\meter}&	std  =  \SI{0,6839}&	CSI	\\\midrule
\cite{MaungMaung2020}&	2020&	WiFi&	exp&	$4$&	$264$&	\SI{112}{\square\meter}&	N&	RF&	&	\SI{1,68}{\meter}&	&	RSSI	\\\midrule
\multirow{3}[4]{*}{\cite{Xiao2020}}&	\multirow{3}[4]{*}{2020}&	WiFi&	exp&	$4$&	$63$&	\SI{75,6}{\square\meter}&	N&	CNN&	&	\SI{1,61}{\meter}&	&	CSI	\\\cmidrule(lr){3-13}
&	&	WiFi&	exp&	$4$&	$N/A$&	\SI{44,8}{\square\meter}&	N&	CNN&	&	\SI{1,11}{\meter}&	& CSI		\\\cmidrule(lr){3-13}
&	&	WiFi&	exp&	$4$&	$N/A$&	\SI{16}{\square\meter}&	N&	CNN&	&	\SI{0,98}{\meter}&	& CSI		\\\midrule
\cite{RN8074}&	2020&	WiFi&	exp&	$4$&	$10$&	\SI{169}{\square\meter}&	Y&	CNN&	&	\SI{0,98}{\meter}&	&	RSSI	\\\midrule
\cite{Vashist2020}&	2020&	WiFi(mmW)&	exp&	$5$&	$34$&	\SI{55}{\square\meter}&	N&	MLP&	Regression&	\SI{0,37}{\meter}&	RMSE  =  \SI{0,84}{\meter}&	RSSI(SNR)	\\\midrule
\cite{RN8008}&	2020&	WiFi&	sim&	$520$&	$993$&	UJIIndoorLoc&	N&	KNN,LR,SVM,RF&	&	&	RMSE = \SI{1,87}{\meter}&	RSSI	\\\midrule
\cite{Ye2020}&	2020&	WiFi&	exp&	$6$&	$112$&	\SI{460}{\square\meter}&	Y&	capsnet&	&	&	Accuracy  =  \SI{0.68}{\meter}&	RSSI	\\\midrule
\cite{Zhang202010773}&	2020&	WiFi&	exp&	$8$&	$133$&	\SI{512}{\square\meter}&	N&	Deep Fuzzy Forest&	&	\SI{1,36}{\meter}&	RMSE  =  \SI{1,79}{\meter}&	RSSI	\\\midrule
\multirow{3}[4]{*}{\cite{Xun2020}}&	\multirow{3}[4]{*}{2020}&	WiFi&	exp&	$1$&	$32$&	\SI{50}{\square\meter}&	N&	DSCP&	&	\SI{1,77}{\meter}&	&	CSI	\\\cmidrule(lr){3-13}
&	&	WiFi&	exp&	$1$&	$24$&	\SI{40}{\square\meter}&	N& DSCP	&	&		\SI{1,16}{\meter}&	& CSI		\\\cmidrule(lr){3-13}
&	&	WiFi&	exp&	$1$&	$66$&	\SI{49}{\square\meter}&	N& DSCP	&	&		\SI{2,54}{\meter}&	& CSI	\\\midrule
\cite{Gao20201426}&	2020&	WiFi&	exp&	$6$&	$50$&	\SI{60}{\square\meter}&	N&	RF&	Bernoulli distribution&	&	RMSE = \SI{2,50}{\meter}&	RSSI	\\\midrule
\multirow{2}[2]{*}{\cite{Zhao2020}}&	\multirow{2}[2]{*}{2020}&	WiFi&	exp&	$25$&	$240$&	\SI{315}{\square\meter}&	N&	RF&	Co-forest&	\SI{2,44}{\meter}&	&	RSSI	\\\cmidrule(lr){3-13}
&	&	WiFi&	exp&	$5$&	$N/A$&	N/A&	N& RF	& Co-forest	&	\SI{4,44}{\meter}&	& RSSI		\\\midrule
\cite{Nabati2020}&	2020&	WiFi&	sim&	$7$&	$1000$&	Rajen Bhatt&	Y&	MLP&	&	&	Accuracy =  94.4\%&	RSSI	\\\midrule
\cite{Qu2020}&	2020&	WiFi&	sim&	$520$&	$993$&	UJIIndoorLoc&	Y&	CNN&	&	&	Accuracy = 88\%&	RSSI	\\\midrule
\cite{RN8121}&	2020&	WiFi&	exp&	$195$&	$300$&	\SI{800}{\square\meter}&	N&	DNN&	HMM&	\SI{1,22}{\meter}&	RMSE = \SI{1,43}{\meter}&	RSSI	\\\midrule
\cite{RN8100}&	2020&	WiFi&	exp&	$3$&	$56$&	\SI{87,75}{\square\meter}&	N&	DNN&	LC&	\SI{0,78}{\meter}&	std =  \SI{1,96}&	CSI	\\\midrule
\cite{RN8037}&	2020&	WiFi&	exp&	$4$&	$236$&	\SI{1148}{\square\meter}&	Y&	BPNN&	GA-PSO&	\SI{0,215}{\meter}&	&	RSSI	\\\midrule
\multirow{2}[2]{*}{\cite{RN8023}}&	\multirow{2}[2]{*}{2020}&	WiFi&	exp&	$10$&	$102$&	\SI{568,4}{\square\meter}&	Y&	LSTM&	LF-D&	\SI{1,48}{\meter}&	&	RSSI	\\\cmidrule(lr){3-13}
&	&	WiFi&	exp&	$30$&	$353$&	\SI{2750}{\square\meter}&	Y&LSTM&	LF-D&	\SI{1,75}{\meter}&	& RSSI		\\\midrule
\multirow{2}[2]{*}{\cite{RN8016}}&	\multirow{2}[2]{*}{2020}&	WiFi&	sim&	$N/A$&	$N/A$&	Cramariuc&	Y&	SEQ2SEQ&	LSTM&	\SI{5,5}{\meter}&	&	RSSI	\\\cmidrule(lr){3-13}
&	&	WiFi&	sim&	$N/A$&	$N/A$&	Cramariuc B&	Y& SEQ2SEQ &	LSTM &	\SI{3,08}{\meter}&	& RSSI		\\\midrule
\multirow{4}[5]{*}{\cite{RN7983}}&	\multirow{4}[5]{*}{2020}&	WiFi&	sim&	$N/A$&	$N/A$&	UAH&	Y&	CNN,LSTM&	&	\SI{4,93}{\meter}&	&	RSSI	\\\cmidrule(lr){3-13}
&	&	WiFi&	sim&	$N/A$&	$N/A$&	CAR&	Y& CNN,LSTM	&	&	\SI{5,4}{\meter}&	& RSSI		\\\cmidrule(lr){3-13}
&	&	WiFi&	sim&	$520$&	$993$&	UJIIndoorLoc&	Y&CNN,LSTM	&	&	\SI{3,2}{\meter}& & RSSI	\\\cmidrule(lr){3-13}
&	&	WiFi&	sim&	$520$&	$993$&	UJIIndoorLoc&	Y&CNN,LSTM	&	&	\SI{4,98}{\meter}& & RSSI	\\\midrule
\cite{RN8069}&	2020&	WiFi&	exp&	$5$&	$22$&	\SI{293}{\square\meter}&	Y&	DNN&	&	&	Accuracy =  95.45\% in \SI{3,65}{\meter} x \SI{3,65}{\meter}&	RSSI	\\\midrule
\multirow{2}[2]{*}{\cite{RN8119}}&	\multirow{2}[2]{*}{2020}&	WiFi&	exp&	$N/A$&	$157$&	\SI{5500}{\square\meter}&	Y&	RNN&	DL&	\SI{3,05}{\meter}&	std =  2.818&	RSSI	\\\cmidrule(lr){3-13}
&	&	WiFi&	sim&	$N/A$&	$157$&	UJIIndoorLoc&	Y& RNN	& DL	&	\SI{4,916}{\meter}&	std =  3.719&	RSSI	\\\midrule
\cite{RN8111}&	2020&	WiFi&	sim&	$54$&	$54$&	\SI{10000}{\square\meter}&	N&	MLP&	&	\SI{3,35}{\meter}&	&	RSSI	\\\midrule
\cite{RN8098}&	2020&	WiFi(mmW)&	exp&	$3$&	$7$&	\SI{25}{\square\meter}&	Y&	DNN&	RESNET&	\SI{0,11}{\meter}&	RMSE =  \SI{0,08}{\meter}&	RSSI(SNR)	\\\midrule
\cite{Chen2020}&	2020&	WiFi&	sim&	$N/A$&	$40$&	UJI Library &	N&	CNN&	SVR&	\SI{2,15}{\meter}&	&	RSSI	\\%\midrule

%% file: Tables/contents2.tex
\cite{RN8625}&	2019&	WiFi&	exp&	$3$&	$30$&	\SI{540}{\square\meter}&	N&	DBN&	Cross entropy&	&	no accuracy reported&	RSSI	\\\midrule
\cite{Schmidt2015}&	2019&	WiFi&	exp&	$2$&	$59$&	\SI{125}{\square\meter}&	Y&	SVM&	&	\SI{0,7}{\meter}&	&	RSSI	\\\midrule
\cite{BelMannoubi2019247}&	2019&	WiFi&	exp&	$N/A$&	$206$&	N/A&	Y&	DNN&	Stacked AutoEncoder&	&	Accuracy =  85\%&	RSSI	\\\midrule
\cite{Yin2019171}&	2019&	WiFi&	exp&	$1$&	$100$&	\SI{100}{\square\meter}&	N&	SVM&	&	\SI{1,909}{\meter}&	std = \SI{0,068}&	RSSI/CSI	\\\midrule
\cite{Naveed2019288}&	2019&	WiFi&	exp&	$N/A$&	$N/A$&	N/A&	N/A&	J4.8,JRIP \& SOM&	CNN&	&	RMSE =  \SI{0,31}{\meter}&	RSSI	\\\midrule
\multirow{2}[2]{*}{\cite{You2019}}&	\multirow{2}[2]{*}{2019}&	WiFi&	exp&	$1$&	$N/A$&	\SI{63}{\square\meter}&	Y&	SVM &	&	& no accuracy reported	&	RSSI	\\\cmidrule(lr){3-13}
& &	WiFi&	exp&	$1$&	$N/A$&	\SI{63}{\square\meter}&	Y&	MLP&	&	& no accuracy reported	&	RSSI	\\\midrule
\cite{Ma2019}&	2019&	WiFi&	exp&	$N/A$&	$N/A$&	N/A&	N/A&	SVM&	&	&	RMSE =  \SI{0,42}{\meter}&	RSSI/CSI	\\\midrule
\cite{Malik2019}&	2019&	WiFi&	exp&	$16$&	$83$&	\SI{305}{\square\meter}&	Y&	DNN&	&	&	Accuracy = \SI{2}{\meter}&	RSSI	\\\midrule
\multirow{2}[2]{*}{\cite{Song2019589}}&	\multirow{2}[2]{*}{2019}&	WiFi&	sim&	$520$&	$993$&	UJIIndoorLoc&	Y&	CNN&	&	&	Accuracy =  95.92\%&	RSSI	\\\cmidrule(lr){3-13}
&	&	WiFi&	sim&	$309$&	$3951$&	Tampere&	Y& CNN	&	&	&	Accuracy =  94.13\%& RSSI		\\\midrule
\cite{Jin2019355}&	2019&	WiFi&	exp&	$6$&	$300$&	\SI{300}{\square\meter}&	N&	MEA-BPNN&	&	\SI{0,72}{\meter}&	&	RSSI	\\\midrule
\cite{RN8321}&	2019&	WiFi&	exp&	$50$&	$N/A$&	N/A&	N/A&	ELM&	&	&	no accuracy reported&	RSSI	\\\midrule
\multirow{2}[2]{*}{\cite{RN8208}}&	\multirow{2}[2]{*}{2019}&	WiFi&	sim&	$520$&	$993$&	UJIIndoorLoc&	Y&	OSELM&	&	\SI{0,7299}{\meter}&	&	RSSI	\\\cmidrule(lr){3-13}
&	&	WiFi&	sim&	$309$&	$1478$&	Tampere&	Y& OSELM	&	&	\SI{0,9274}{\meter}&	& RSSI		\\\midrule
\cite{RN8240}&	2019&	WiFi&	exp&	$256$&	$74$&	\SI{1664}{\square\meter}&	Y&	CNN&	&	&	Accuracy =  95.4\% in \SI{4}{\meter}&	RSSI	\\\midrule
\cite{RN8415}&	2019&	WiFi&	exp&	$54$&	$180$&	\SI{1209}{\square\meter}&	Y&	RDF&	&	&	Accuracy =  89\% at room level&	RSSI	\\\midrule
\cite{RN8169}&	2019&	WiFi&	exp&	$256$&	$74$&	\SI{300}{\square\meter}&	Y&	CNN&	&	\SI{1,46}{\meter}&	Accuracy =  94\% std = 2.24&	RSSI	\\\midrule
\cite{RN8409}&	2019&	WiFi&	exp&	$4$&	$42$&	\SI{80}{\square\meter}&	N&	RBF&	LM&	\SI{1,421}{\meter}&	RMSE = \SI{1,459}{\meter}&	RSSI	\\\midrule
\cite{RN8301}&	2019&	WiFi&	exp&	$N/A$&	$300$&	\SI{302}{\square\meter}&	Y&	SVM&	&	\SI{4,6}{\meter}&	&	RSSI	\\\midrule
\cite{RN8235}&	2019&	WiFi&	exp&	$5$&	$10$&	N/A&	N&	RF&	&	&	Accuracy = 97.5\% in \SI{2}{\meter}&	RSSI	\\\midrule
\cite{RN8381}&	2019&	WiFi&	exp&	$8$&	$107$&	\SI{512}{\square\meter}&	Y&	K-ELM&	&	&	RMSE =  \SI{1,7123}{\meter} std = \SI{2,418}&	RSSI	\\\midrule
\cite{RN8182}&	2019&	WiFi&	exp&	$9$&	$96$&	\SI{560}{\square\meter}&	Y&	QKMMCC-BG&	&	&	average =  \SI{0,76}{\meter} &	RSSI	\\\midrule
\multirow{2}[2]{*}{\cite{RN8310}}&	\multirow{2}[2]{*}{2019}&	WiFi&	sim&	$520$&	$993$&	UJIIndoorLoc&	Y&	RNN&	&	&	Accuracy = 96.45\% building  87.41\% floor&	RSSI	\\\cmidrule(lr){3-13}
&	&	WiFi&	exp&	$7$&	$N/A$&	4 Rooms&	Y& RNN	&	&	&	Accuracy = 95.8\% std =  \SI{0.60}& RSSI		\\\midrule
\multirow{2}[2]{*}{\cite{RN8134}}&	\multirow{2}[2]{*}{2019}&	WiFi&	sim&	$520$&	$993$&	UJIIndoorLoc&	Y&	RNN&	&	\SI{4,2}{\meter}&	std =  \SI{3,2}&	RSSI	\\\cmidrule(lr){3-13}
&	&WiFi	&	exp&	$6$&	$365$&	\SI{336}{\square\meter}&	Y&	RNN&	&	\SI{0,75}{\meter}&	std =  \SI{0,64}&	RSSI	\\\midrule
\cite{RN8280}&	2019&	WiFi&	exp&	$9$&	$261$&	\SI{300}{\square\meter}&	N&	BPNN &	&	\SI{2,7}{\meter}&	Accuracy =  90\%&	RSSI	\\\midrule
\cite{RN8295}&	2019&	WiFi&	exp&	$8$&	$66$&	\SI{736}{\square\meter}&	Y&	SDA&	&	\SI{3,7}{\meter}&	Accuracy =  84\%&	RSSI	\\\midrule
\multirow{2}[2]{*}{\cite{RN8148}}&	\multirow{2}[2]{*}{2019}&	WiFi&	exp&	$1$&	$42$&	\SI{50}{\square\meter}&	N&	CNN&	&	\SI{0,46}{\meter}&	without obstacles& RSSI		\\\cmidrule(lr){3-13}
&	&	WiFi&	exp&	$1$&	$42$&	\SI{50}{\square\meter}&	N&	CNN&	&	\SI{1,11}{\meter}&	with some obstacles& RSSI		\\\midrule
\multirow{4}[5]{*}{\cite{RN8345}}&	\multirow{4}[5]{*}{2019}&	WiFi&	exp&	$1$&	$15$&	\SI{20}{\square\meter}&	Y&	MLP&	&	\SI{1,42}{\meter}&	&	RSSI	\\\cmidrule(lr){3-13}
&	&	WiFi&	exp&	$1$&	$15$&	\SI{20}{\square\meter}&	Y&	CNN&	&	\SI{1,67}{\meter}&	& RSSI		\\\cmidrule(lr){3-13}
&	&	WiFi&	exp&	$1$&	$15$&	\SI{14,4}{\square\meter}&	N&	MLP&	&	\SI{1,43}{\meter}&	& RSSI		\\\cmidrule(lr){3-13}
&	&	WiFi&	exp&	$1$&	$15$&	\SI{14,4}{\square\meter}&	N&	CNN&	&	\SI{1,51}{\meter}&	& RSSI	\\\midrule
\cite{RN8149}&	2019&	WiFi&	exp&	$258$&	$9$&	\SI{125}{\square\meter}&	Y&	CNN&	&	\SI{3,91}{\meter}&	Accuracy =  84\%&	RSSI	\\\midrule
\cite{RN8373}&	2019&	WiFi&	sim&	$N/A$&	$N/A$&	N/A&	Y&	BPNN&	ACO&	&	Accuracy =  91.4\% &	RSSI	\\\midrule
\cite{RN8205}&	2019&	WiFi&	sim&	$520$&	$993$&	UJIIndoorLoc&	Y&	CNN, GRP&	&	\SI{1}{\meter}&	\SI{2,633}{\meter} with knn&	RSSI	\\\midrule
\cite{RN8217}&	2019&	WiFi&	exp&	$1$&	$25$&	\SI{26,4}{\square\meter}&	N&	BPNN &	PCA-PD&	\SI{1,4223}{\meter}&	std =  \SI{1,1511}&	CSI	\\\midrule

%% file: Tables/contents3.tex
\multirow{3}[4]{*}{\cite{RN8203}}&	\multirow{3}[4]{*}{2019}&	WiFi&	exp&	$N/A$&	$20$&	\SI{1200}{\square\meter}&	Y&	MLP&	SDAE&	\SI{3,05}{\meter}&	1day&	RSSI	\\\cmidrule(lr){3-13}
&	&WiFi	&	exp&	$N/A$&	$57$&	\SI{2400}{\square\meter}&	Y&	MLP&	SDAE&	\SI{3,39}{\meter}&	2days&	RSSI	\\\cmidrule(lr){3-13}
&	&WiFi	&	sim&	$520$&	$993$&	UJIIndoorLoc&	Y&	MLP&	SDAE&	\SI{5,64}{\meter}&	10days&	RSSI	\\\midrule
\cite{RN8323}&	2019&	WiFi&	sim&	$520$&	$993$&	UJIIndoorLoc&	Y&	VAE&	&	&	RMSE =  \SI{4,65}{\meter}&	RSSI	\\\midrule
\multirow{2}[2]{*}{\cite{RN8578}}&	\multirow{2}[2]{*}{2019}&	WiFi&	exp&	$6$&	$49$&	\SI{1600}{\square\meter}&	Y&	DNN&	&	\SI{0,952}{\meter}&	Open Doors&	RSSI	\\\cmidrule(lr){3-13}
&	&Wifi	&	exp&	$6$&	$49$&	\SI{1600}{\square\meter}&	Y&	DNN&	&	\SI{1,258}{\meter}&	Closed Doors& RSSI		\\\midrule
\cite{RN8191}&	2019&	WiFi&	exp&	$4$&	$228$&	\SI{1200}{\square\meter}&	Y&	ANN&	&	\SI{1,2244}{\meter}&	&	RSSI \\\midrule
\cite{RN8391}&	2019&	WiFi&	exp&	$7$&	$25$&	\SI{1728}{\square\meter}&	N&	RNN&	LSTM&	\SI{1,0538}{\meter}&	std = \SI{0,8856}&	RSSI	\\\midrule
\cite{RN8274}&	2019&	WiFi&	exp&	$15$&	$71$&	\SI{4000}{\square\meter}&	Y&	NN&	GA&	\SI{3,47}{\meter}&	&	RSSI	\\\midrule
\cite{RN8350}&	2019&	WiFi&	exp&	$4$&	$50$&	\SI{1100}{\square\meter}&	Y&	BGM&	&	\SI{2,9}{\meter}&	&	RSSI	\\\midrule
\cite{RN8358}&	2019&	WiFi&	exp&	$122$&	$48$&	\SI{629}{\square\meter}&	Y&	DNN&	&	\SI{2,64}{\meter}&	&	RSSI	\\%\midrule

%% file: 04-results.tex
\section{Results}
This section presents the results obtained. The complete diagram of the articles obtained can be seen in Figure~\ref{fig:foo}.

Table \ref{tabla:review} displays the summary with the main features of the $64$ articles reviewed in detail. Please note that:

\begin{itemize}
\item Features not explained in the articles appear as $N/A$.%Los resultados no explicados en los artículos se han dejado en blanco.
\item Articles that include different experiments and/or simulations are shown grouped together.
\item Articles that do not display a metric clear enough, are marked in the column oError  (Other Errors).
\item Articles that are based on or use algorithms different from the main one, are marked in the column sAlg (Secondary Algorithm)
\end{itemize}

%% file: 05-discussion.tex
\section{Discussion}
In this section we will analyze the results from several points of view: algorithms used, types of signals used, number of APs and reference points used, metrics, type of experimentation, most commonly used radio maps and the use or not of rooms on these maps.

\subsection{Algorithms}
The most commonly used algorithms are those based on artificial neural networks (ANN). Specifically we see up to 39 articles (around $61\%$ of total analysed works) that use this machine learning model or any of its variants (CNN, DNN, BPNN, capsnet, RNN, DSCP, MLP, QKMMCC, NN). The fluctuating signal type of WiFi makes it a good algorithm because it is especially suitable for nonlinear functions, such as WiFi signals. In addition, we see how the best result in the analyzed papers (mean error of \SI{0.11}{\meter}) is obtained by using precisely a Deep Neural network to process data from WiFi mmWave signal \cite{RN8098}.

We also find different articles that focus on combining different algorithms to choose the one that best results in a given case \cite{Roy2021,RN8008}, others that focus on processing data collected from APs \cite{Nabati2020,Qu2020,RN8023,RN8016} and others \cite{RN8037,RN8119} that rely on applying a double algorithm (one to approximate the location and one to further specify it).

\subsection{Types of WiFi signal parameter used}
Regarding the signal information used for positioning, most works are based on received signal strength (RSSI). That is because this signal parameter is easily accessible from any WiFi device (including smartphones and wearable devices). In general, the results obtained with RSSI are around a few meters, although range from almost one meter to several meters. Looking at the data in the table it is easy to see that when smaller and more reference points are used the results improve. Using spaces without rooms also has an impact on the results, that tend to be better. This is explained both, by the effect of signal loss when passing through walls, and the multipath effect, as explained in \cite{Obeidat2018}. It is important to note that although  the error in RSSI-based methods is usually higher, paper~\cite{RN8037} obtains the second best result among those analyzed \SI{0.22}{\meter}.

On the other hand, there are two studies, \cite{Xun2020,RN8100}, that use channel state information (CSI) from a WiFi signal and generally with better results in accuracy than those obtained with RSSI. CSI is not widely used because the channel state information is not easy to obtain and requires specific network cards and modifications on the original \textit{firmware} ~\cite{Xun2020} (i.e., it cannot be used in smartphones).

Finally, we see that the SNR parameter is starting to be used, in particular in combination with those WiFi networks that use \emph{mmWave}~\cite{RN8098,Vashist2020} instead of the classic networks that broadcast on traditional frequency channels, \SI{2.4}{\giga\hertz} and \SI{5}{\giga\hertz}. We also see better positioning accuracy with these technologies, with which the best and the third best results are obtained.

\subsection{Metrics}
In order to compare works among themselves, a common metric is needed. Most jobs report their results with the absolute positioning error mean (euclidean distance between two points), \emph{Mean Error} on the table. Among them, most also report the root mean squared error, \emph{RMSE} on the table. Other metrics used are the mean squared error (\emph{MSE}) and the median error (\emph{MedianError}). On the other hand, 14 papers report the error of success in a certain precision based in percentage \cite{Roy2021,Ye2020,Nabati2020,Qu2020,RN8069,BelMannoubi2019247,Malik2019,Song2019589,RN8235,RN8240,RN8415,RN8373,RN8169,RN8310}. Unfortunately, three articles do not show results clearly or are based on items different from those analyzed in this review \cite{Hsu2018132,You2019,RN8321}.

\subsection{Experimental and simulated results}

Specifically, we found 27 that present results from simulations and 63 that present empirical results (note that there are articles that perform several experiments and/or simulations). The greater presence of empirical work  may be due to an improvement in the APs and mobile devices used today in comparison to those that were used in the different radio maps used in the simulations or by a better situation of the different devices in the actual experiments (optimize the position to force a better result). Three papers apply the same method in both, simulation and experimentation \cite{RN8119,RN8310,RN8134}. In two of them, \cite{RN8119,RN8134}, there is an improvement in the results obtained with empirical experimentation.  However, this improvement can be due, among other reasons, to the greater number of reference points used to create the WiFi radio map in relation to the space of the experiment compared to the maps used in simulation.

\subsection{Most used radio-maps}
In the list, the most commonly used public signal map is the UJIIndoorLoc~\cite{7275492} in all its variants (different buildings). It appears in 14 papers and is clearly the most important, since the rest of the maps found appear only once.

The $35\%$ of experiments reported in 2020 were performed on public databases, whereas only $18\%$ of experiments in 2019 had a similar experimental setup based on external datasets. Having datasets enables researchers to provide useful results as well as performing experiments under a pandemic situation. Others radio-maps used are IPIN2016~\cite{8115940}, UTSIndoorLoc~\cite{8792196}, JUIndoorLoc~\cite{Roy2019},Rajen Bhatt~\cite{Rohra2017} and Cramariuc~\cite{7533846}.

On the other hand, there is  a great diversity of areas in which experiments take place. Most have been done in university facilities. Specifically, with less than \SI{100}{\square\meter} we have $16$ articles, $14$ between \SI{100}{\square\meter} and \SI{500}{\square\meter}, $8$ between \SI{500}{\square\meter} and \SI{1000}{\square\meter} and $13$ higher than \SI{1000}{\square\meter}. It should be noted that there are articles with more than one experiment and that are therefore part of several groups.
Another important aspect of WiFi fingerprinting Radio Maps is whether they are used in spaces without rooms or with rooms. Specifically we see that 57 of the radio maps used have rooms in their experimentation space and there are 30 that are in spaces without any rooms (spaces without walls).
Finally, it should be noted that the vast majority of jobs are carried out with the premise of obtaining the best results and do not perform the experiments in an everyday environment.

%% file: 06-conclusions.tex
\section{Conclusions}
This work has shown a systematic review of the current state in the application of deep learning algorithms applied to indoor positioning using WiFi. Information from 64 articles between 2019 and 2021 has been extracted and analysed.

This study shows a tendency to use neural network-based algorithms for positioning based on WiFi, and UJIIndoorLoc as the most used public signal map for most simulations. However, it should be noted that the studies with the best results in terms of accuracy are those carried out experimentally. However, many of these results are obtained in small workspaces with a lot of reference points and/or APs.

We can also note the predominant use of the RSSI of WiFi signals, although the studies that focus on the CSI are very promising, since they are the ones that have obtained the best accuracy.

For future work, we plan to extend the review to a longer period of time to detect whether there are any remarkable trends that have occurred in recent years and to extend the analysis to the computing time required to obtain the results, since in certain circumstances this can be a clear reason for choosing one technology or another. Attention should also be paid to the treatment of signals before being processed by deep learning algorithms, as it has been found that some studies do perform previous pre-processing work but others focus only on the output result of the algorithm.

%% file: 07-acknowledgments.tex
\section*{Acknowledgment}

Authors wants to thank the Spanish Network of Excellence REPNIN+ (MICINN, ref. TEC2017-90808-REDT). J. Torres-Sospedra acknowledges funding from Torres Quevedo programme (PTQ2018-009981).